\documentclass[sigchi]{acmart}

\makeatletter
\def\mdseries@tt{m}
\makeatother

\usepackage{booktabs} 
\usepackage{xspace}
\usepackage{url}
\usepackage{graphicx}
\usepackage{xcolor}
\usepackage[frozencache]{minted}
\usepackage{verbatim}
\usepackage{subcaption}
\usepackage{cuted}
\usepackage{capt-of}
\usepackage[normalem]{ulem}
\usepackage{microtype}

\copyrightyear{2019}
\acmYear{2019}
\setcopyright{acmcopyright}
\acmConference[IUI '19]{24th International Conference on Intelligent User Interfaces}{March 17--20, 2019}{Marina del Ray, CA, USA}
\acmBooktitle{24th International Conference on Intelligent User Interfaces (IUI '19), March 17--20, 2019, Marina del Ray, CA, USA}
\acmPrice{15.00}
\acmDOI{10.1145/3301275.3302314}
\acmISBN{978-1-4503-6272-6/19/03}

\newcommand{\eg}{\textit{e.g.}\xspace}
\newcommand{\ie}{\textit{i.e.}\xspace}

\BeforeBeginEnvironment{minted}{\vspace{0.1cm}}
\AfterEndEnvironment{minted}{\vspace{0.1cm}}

\begin{document}
\title{\textls[-50]{SAM: A Modular Framework for Self-Adapting Web Menus}}

\author{Camille Gobert\textsuperscript{1} \hspace{2cm} Kashyap Todi\textsuperscript{1} \hspace{2cm} Gilles Bailly\textsuperscript{2} \hspace{2cm} Antti Oulasvirta\textsuperscript{1}}
\affiliation{%
  \institution{camille.gobert@ens.fr\hspace{1cm}  kashyap.todi@gmail.com\hspace{1cm}  gilles.bailly@upmc.fr\hspace{1cm}  antti.oulasvirta@aalto.fi}
  \institution{\textsuperscript{1}Department of Communication and Networking, Aalto University\hspace{0.5cm}\textsuperscript{2}Sorbonne Universit\'e, CNRS, ISIR}
  }

\renewcommand{\shortauthors}{Gobert, C. et al.}

\begin{abstract}
This paper presents SAM, a modular and extensible JavaScript framework for \underline{s}elf-\underline{a}dapting \underline{m}enus on webpages.
SAM allows control of two elementary aspects for adapting web menus:
(1) the \emph{target policy}, which assigns scores to menu items for adaptation, and
(2) the \emph{adaptation style}, which specifies how they are adapted on display.
By decoupling  them,
SAM enables the exploration of different combinations independently.
Several policies from literature are readily implemented,
and paired with adaptation styles such as reordering and highlighting.
The process---including user data logging---is local,
offering privacy benefits and eliminating the need for server-side modifications.
Researchers can use SAM to experiment adaptation policies and styles, and benchmark techniques in an ecological setting with real webpages.
Practitioners can make websites self-adapting,
and end-users can dynamically personalise typically static web menus.
\end{abstract}
%
%

\ccsdesc[500]{Human-centered computing~Human computer interaction (HCI)}
\ccsdesc[500]{Human-centered computing~User interface design}
%
%



\maketitle


\section{Introduction}
Several decades of work on adaptive menus has shown empirical evidence that they can improve the usability of complex interfaces  \cite{sears1994split,tsandilas2007bubbling,Cockburn2007,findlater2009ephemeral,bailly2017visual}.
However, most of these techniques have not been deployed or even released \cite{bailly2017visual}.
Technical and practical reasons can be identified.
First, to improve adaptive interfaces, it is crucial to isolate and understand the different aspects of the adaptation process  \cite{gajos2006exploring}.
Second, implementing adaptive menus that work outside prototype systems remains a challenging task.
Most graphical toolkits (\eg Java Swing, Qt) only provide limited support for customisation \cite{bailly2017visual}.
Our goal is to support effective (re)use of theoretical and technical knowledge in adaptive menus to facilitate the transfer of this technology \cite{Sutcliffe:2000:EUR:353485.353488}.

UI toolkits and frameworks have facilitated the implementation of software interfaces \cite{Myers:1995:UIS:200968.200971}.
Similarly, we aim to support implementation and adoption of adaptive menus and interfaces.
To this end, we present SAM, an open-source, modular and extendable JavaScript framework for the research and deployment of \underline{s}elf-\underline{a}dapting \underline{m}enus on regular web pages and applications.
SAM offers logging capability for modern browsers and permits full control of two elementary aspects of an adaptive menu:
(1) a \textbf{target policy}, which determines relative importance of items and groups found within a menu, based on a user's interaction history, and
(2) an \textbf{adaptation style}, which specifies the visual changes made while adapting menu items (\ie look and feel).
By decoupling the two, it is possible to independently explore different combinations.
To facilitate this, SAM includes several readily implemented adapted menus from the literature (\eg \cite{mitchell1989dynamic,tsandilas2007bubbling,Cockburn2007,fitchett2012accessrank}).
Finally, a key aspect of SAM's design is to ensure privacy:
the adaptation process---including user data logging and computation---is entirely local,
and requires no server-side storage or modifications.

SAM targets researchers, practitioners, and end-users.
Researchers can use SAM to experiment adaptation policies and styles, and compare with previously published ones, in ecological settings (with real webpages).
Practitioners can make their websites self-adapting with minor modifications.
Finally, end-users can dynamically personalise both the policy and the style of web menus which normally are static.

Our main contribution is the design and implementation of SAM, an open-source framework promoting ecological validity, replicability, and transfer of research on adaptive menus.
%
To demonstrate the capabilities and usage of the framework,
we implemented six policies and four styles issued from literature, leading to 24 different adaptive menus.
We describe the code required to create a new design and integrate it in an existing web page.

\section{Related Work}
This work is situated within two specific areas: (1) menu adaptation techniques and (2) automatic adaptation of webpages.
We provide a brief overview of each as a precursor and motivation to SAM's design.

\subsection{Menu Adaptation}

Several menu adaptation styles have been proposed \cite{bouzit2016design,bailly2017visual,Vanderdonckt18}.
For instance, frequency-ordered or folded menus move the most frequent items to the top of the menu \cite{bailly2017visual,lee2004quantitative}.
However this approach does not maintain any visual consistency.
To avoid this problem, split menus \cite{sears1994split} duplicate the 2--5 most frequent items on the top of the menu so that the bottom part of the menu remains unchanged.
This solution has been transposed from a single menu to the whole menu system \cite{liu2018bigfile}.
Other approaches increase the saliency of some items by manipulating the size \cite{Cockburn2007}, the transparency \cite{bailly2017visual}, the background colour \cite{tsandilas2007bubbling}, or the delay of apparition \cite{findlater2009ephemeral,lee2004quantitative}  of the items.
Except \cite{gajos2006exploring}, which studied three different menu adaptation styles and two policies, previous works generally rely on one ``simple'' target policy (\ie item frequency) and do not study the combination of one or several styles.
Moreover, these implementations and studies are restricted to specifically-developed software applications.
This reduces the potential for adoption and replicability.

\subsection{Adaptation of Webpages}

AI techniques have been used to mine user logs for webpage adaptation \cite{perkowitz_adaptive_1997}.
PageGather \cite{perkowitz_towards_1999} uses this to automatically create index pages by identifying candidate link sets on websites.
However, all data logging and computation is done on the server-side,
and the adaptation is constricted to creating new index pages.
PageTailor \cite{bila_pagetailor:_2007} supports client-side customisation of webpages by the user, and saves the results for revisits.
Similarly, \cite{tsandilas_user-controlled_2003} supports end-user link adaptation on webpages for better information discovery in long menus.
These works lack policies for automatic adaptation.
\cite{kurniawan_personalising_2006} 
automatically adapts pages based on styling issues related to accessibility, but does not continually adapt them based on usage and policies.
\cite{todi2018familiarisation} discusses automatic client-side web page layout adaptation to make them familiar to users based on usage histories.
The adaptation is however limited to changes in positions.
\cite{leiva_restyling_2011} develops an adaptive engine that took user's touch input to adapt element style based on usage.
Similarly to our work, this client-side engine applies a usage-based policy to adapt styles using web technologies.
However, the system does not decouple the different aspects of the adaptation, and thus hinders the exploration of different combinations of policies and styles, or further expandability---which are the key goals of our framework.

\section{SAM: The Framework}

\begin{figure}
\includegraphics[width=0.9\columnwidth]{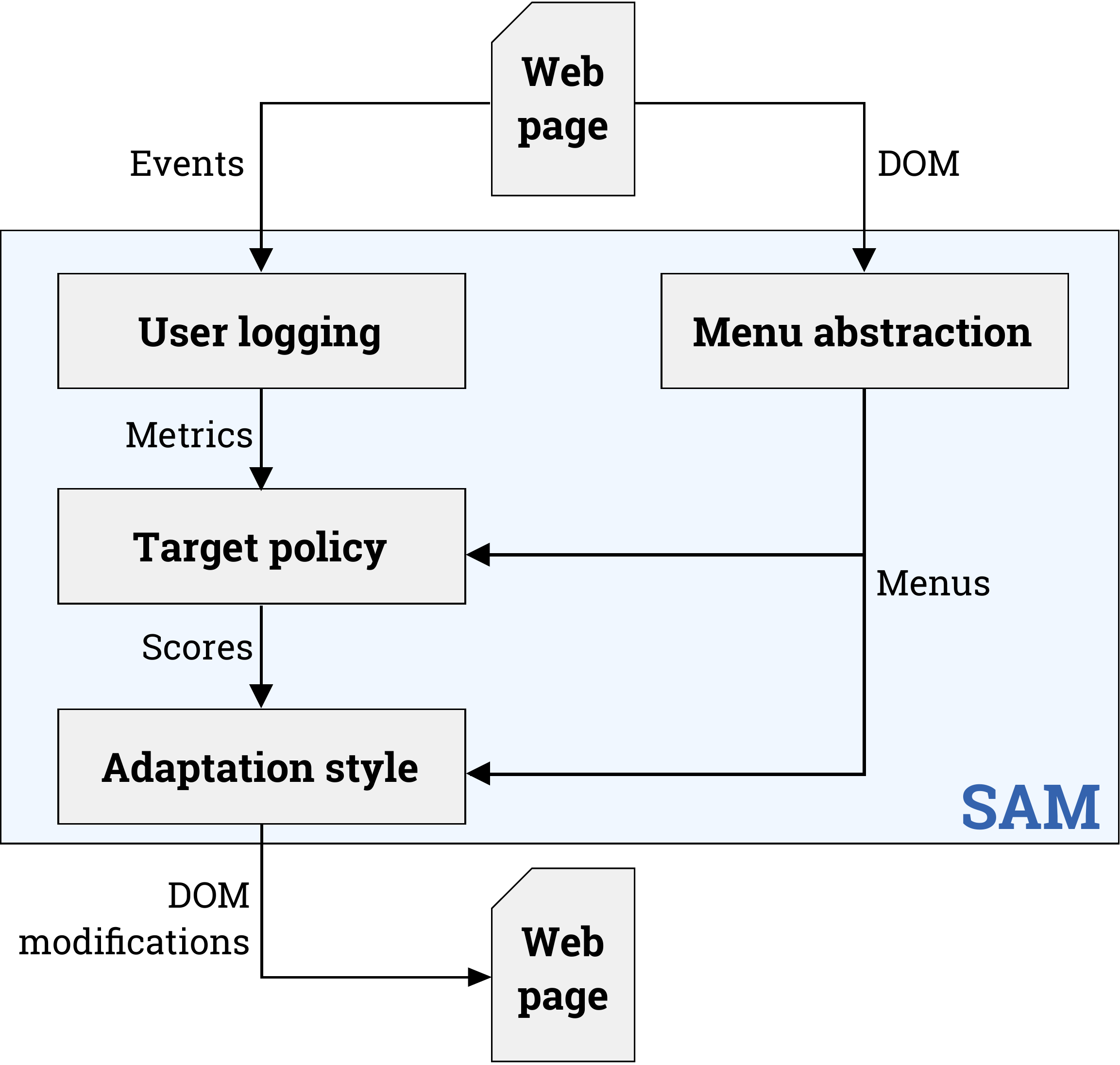}
\caption{Overview of the interactions between the core modules of the SAM framework.}
\label{fig:implem-scheme}
\end{figure}

SAM is a client-side framework (\autoref{fig:implem-scheme}).
By separating out the underlying modules, we enable flexibility and control over the adaptation process.
In this section, we provide an overview of each module. The implementation is described in the following section.

\subsection{Menu Abstraction}
Webpage menus are defined as sets of elements,
whose types and hierarchy tend to differ from one website to another.
To address this diversity,
SAM builds a single common abstraction of the structure of a menu.
This representation is necessary to consistently capture menu usage and adapt them across very different websites.

\subsection{User Logging and Metrics}
\begin{figure*}
  \captionsetup{justification=centering}
  \begin{subfigure}[t]{0.35\textwidth}
    \vspace{0.27cm}
    \centering
    \footnotesize
    \begin{tabular}{rrr}
    \toprule
    \textbf{Item label} & \textbf{Nb. clicks} & \textbf{Duration (s)} \\
    \midrule
    Main page           & 2  & 74 \\
    \textcolor{gray}{Contents}            & \textcolor{gray}{0}  & \textcolor{gray}{0} \\
    Featured contents   & 6  & 29 \\
    \textcolor{gray}{Current events}      & \textcolor{gray}{0}  & \textcolor{gray}{0} \\
    Random article      & 10 & --- \\
    \textcolor{gray}{Donate to Wikipedia} & \textcolor{gray}{0}  & \textcolor{gray}{0} \\
    \textcolor{gray}{Wikipedia store}     & \textcolor{gray}{0}  & \textcolor{gray}{0} \\
    \bottomrule
    \end{tabular}
    \caption{Interaction history.}
    \label{subfig:wiki-interaction-history}
  \end{subfigure}
  \begin{subfigure}[t]{0.20\textwidth}
  	\vspace{0pt}
    \centering
    \includegraphics[width=0.8\linewidth, trim={0 0 0 3.95cm}, clip]
    	{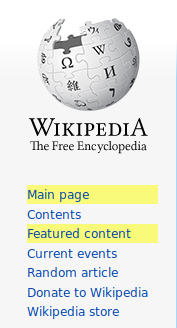}
    \caption{Highlighting\\by Visit duration.}
    \label{subfig:wiki-highlighting-visitduration}
  \end{subfigure}
  \begin{subfigure}[t]{0.20\textwidth}
    \vspace{0pt}
    \centering
    \includegraphics[width=0.8\linewidth, trim={0 0 0 3.95cm}, clip]
    	{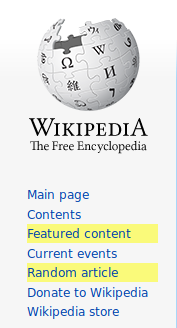}
    \caption{Highlighting\\by Click frequency.}
    \label{subfig:wiki-highlighting-clickfreq}
  \end{subfigure}
  \begin{subfigure}[t]{0.20\textwidth}
    \vspace{0pt}
    \centering
    \includegraphics[width=0.8\linewidth, trim={0 0 0 3.95cm}, clip]
    	{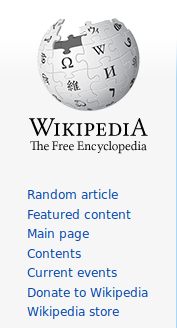}
    \caption{Item reordering\\by Click frequency.}
    \label{subfig:wiki-reordering-clickfreq}
  \end{subfigure}
  \caption{Various adaptive menus produced by SAM on Wikipedia, given a single interaction history (a).\\Changing the target policy ($\text{b} \rightarrow \text{c}$) or the adaptation style ($\text{c} \rightarrow \text{d}$) both result in different adaptive menus.}
  \label{fig:wiki-examples}
\end{figure*}

SAM captures a user's interaction history with menus,
and stores this locally in a database.
Currently, SAM records mouse clicks and time spent on each page.
To extend logging capabilities,
any event that can be captured on a webpage (\eg cursor movements, eye tracking),
can be logged by SAM, and used for further computations.
SAM then computes usage metrics (\eg click frequency or page visits) from user's interaction history to update the selected target policy.
While this history may include erroneous events (such as clicking the wrong item), they eventually become neglected as the system captures more legitimate events and self-corrects itself.
\autoref{subfig:wiki-interaction-history} shows some logged data for an example scenario.

\subsection{Target Policies}
Scores are assigned to all menu items (and/or groups) based on a target policy.
While new policies can be easily integrated into SAM, we include a readily-available set of policies, adapted from literature:
\begin{enumerate}
\item \textbf{Item clicks frequency:} Items are assigned normalised scores based on the number of clicks \cite{mitchell1989dynamic,sears1994split}.
\item \textbf{Page visits duration:} Items are assigned normalised scores based on the total time spent on the page they link to.
\item \textbf{Page visits frequency:} 
Items linking to pages visited more frequently are assigned higher scores \cite{todi2018familiarisation}.
\item \textbf{Page visits recency:} Items linking to pages visited more recently are assigned higher scores \cite{todi2018familiarisation}.
\item \textbf{Serial-position curve:} Items are assigned normalised scores combining frequency, recency, and primacy \nolinebreak\cite{todi2018familiarisation}.
\item \textbf{AccessRank:} Items are scored according to the AccessRank algorithm \cite{fitchett2012accessrank}, which includes recency, frequency, temporal clustering, and time of day as factors as well as a component for stability.
\end{enumerate}

\subsection{Adaptation Styles}
SAM uses the score of each item (target policy) to apply the corresponding adaptation style to items.
The number ($N$) of items to be adapted can either be specified as a fixed number, or as a function of the menu size.
Currently, SAM includes the following styles:
\begin{enumerate}
\item \textbf{Highlighting:} $N$ items are highlighted using a contrasting background colour\footnote{ Or any other effect which can be applied with CSS.} \cite{tsandilas2005empirical}.
\item \textbf{Item reordering:} Selected $N$ items are moved to the top of the (sub-)menu \cite{mitchell1989dynamic}.
\item \textbf{Group reordering:} Selected $N$ entire groups are moved to the top of the menu.
\item \textbf{Folding:} Items with low scores are truncated from the menu \cite{bailly2017visual}.
\end{enumerate}
SAM can include additional styles (\eg transparency, size, font, borders, positions, etc.) and can combine them to form composite styles.


\section{Implementation}

SAM is implemented in Typescript\footnote{ \url{https://www.typescriptlang.org}}, a typed scripting language which is compiled to plain JavaScript.
The only external dependency is jQuery\footnote{\url{https://jquery.com}; the \textit{slim} version suffices for SAM.}, which is used to facilitate Document Object Model\footnote{\url{https://www.w3.org/DO/}} (DOM) manipulations.
To make SAM extendible, it has been split into multiples modules with different responsibilities (Figure \ref{fig:implem-scheme}).

\subsubsection{Menu abstraction}
Each element of adaptive menus (\eg item, group) is associated with a node in the DOM.
For a menu to adapt, a set of jQuery selectors must be provided to SAM, in order to fetch the nodes which form the structure of the menu in the webpage.
Each element is then given a unique identifier, used to track them across pages and sessions.
The identifier is determined by the node tag, \texttt{id} attribute, position among its siblings, and those of its ancestors.


\subsubsection{Interaction Logging}
The logging of all user interactions has been split between three modules to keep the code easily understandable and simple to extend:
The \emph{data logger} logs any event they catch in the database. It currently logs all \texttt{click} events fired on menu items, and all page visits (using \linebreak\texttt{beforeunload} events).
%
The \emph{database} serialises its content into the Local Storage of the browser, and un-serializes it on each page load.
This choice allows to easily switch to another form of persistent storage (\eg IndexedDB), or to send data to a remote server (\eg for an online study).
The \emph{data analyser} 
improves performances, by only recomputing metrics if the database content has been updated since the last computation, and using a cached version otherwise.

\subsubsection{Target Policy}
The target policy uses the output of the data analyser and abstract menus content (\autoref{fig:implem-scheme}) to compute the scores for each item or/and group and sort them.



\subsubsection{Adaptation Style}
The adaptation style takes the output of the aforementioned policy and a list of abstracted menus to modify the DOM and apply the desired effect to the target items or/and groups.
To be compatible with SAM, each style must implement two methods: one to apply the effect of the style, and one to cancel it.


\subsubsection{Privacy}
User data, target policy, and adaptation styles are stored, computed, and applied locally, on the user's browser.

\subsubsection{Scalability}

SAM can adapt web menus (desktop or mobile) of all sizes, with or without groups, with hundreds of items.
The main threshold is the complexity of the target policy and the adaptation style---which can run arbitrarily costly computations.
However, modern web browsers heavily optimise JavaScript code they run, and any combination of currently implemented policy and styles can smoothly adapt menus with dozens of groups and hundreds of items in a few milliseconds\footnote{In our tests, SAM computations on \url{wikipedia.org} took 0.077s (average over 10 page loads).}.
Furthermore, although the typical Local Storage of a browser has limited capacity ($\sim$\,5--10 MB per domain), the database of SAM can still store 2,000 to 10,000 visits and clicks before exceeding the available space.
This limit could be easily bypassed by using IndexedDB instead of Local Storage.

\section{Usage and Extension}

We release SAM as an open-source framework at \textcolor{blue}{\url{https://github.com/aalto-ui/sam}}.
It can be adopted and/or extended by different categories of users, based on their objectives.

\subsubsection{Developers}
Developers can edit the sources of a website to include SAM, and turn their static menus into adaptive menus.
They first include the JavaScript library and CSS file in the HTML sources of any page with menus to adapt:

\begin{minted}[fontsize=\footnotesize]{html}
<link rel="stylesheet" href="css/sam.css"></link>
<script type="text/javascript" src="js/sam.js"></script>
\end{minted}

They then initialise the framework by calling the static builder method \texttt{fromSelectors} on the global \texttt{SAM} object---which is exposed in the global scope by \texttt{sam.js}. This step requires the DOM to be fully loaded:


\begin{minted}[fontsize=\footnotesize]{js}
$(document).ready(() => {
  SAM.fromSelectors(".menu", ".group", ".item");
});
\end{minted}



\subsubsection{Researchers}
Researchers can extend SAM with new policies and styles.
To do so, they must implement the \texttt{TargetPolicy} or \texttt{AdaptationStyle} interfaces, add the new class to the right internal array, and recompile the framework (by running \texttt{grunt}\footnote{ \url{https://gruntjs.com/}}).
For example, one could add a \texttt{Magnify} style which increases the font size of the 3 elements with the highest scores. By adding a CSS rule which increases the font size of all elements with class \texttt{sam-magnified}, the following snippet is sufficient to implement this style:


\begin{minted}[fontsize=\footnotesize]{ts}
public class Magnify implements AdaptationStyle {
  readonly name = "Magnify";
  readonly    N = 3; // Number of items to select

  apply(menuManager, policy, dataManager) {
    // Select top N items ranked by the policy
    let items = policy.getSortedItems(menuManager, dataManager)
                      .slice(0, N);
    for (let item of items) {
      item.node.addClass("sam-magnified");
  } }

  cancel() {
    $(".sam-magnified").removeClass("sam-magnified");
} }
\end{minted}



For convenience, abstract classes with partial implementations are provided. The following snippet illustrates how this allows to add a new target policy to SAM by implementing only one method:

\begin{minted}[fontsize=\footnotesize]{ts}
public class ExamplePolicy extends DefaultTargetPolicy {
  readonly name = "ExamplePolicy";

  getSortedItemsWithScores(menuManager, dataManager) {
    let items = menuManager.getAllItems();
    let itemsWithScores = items.map((item) => {
      return {item: item, score: 0};
    });

    // --- Code to compute item scores ---

    return itemsWithScores;
} }

\end{minted}

\subsubsection{End-users}
End-users can use SAM to adapt any web menu.
This is achieved by injecting the SAM JavaScript library, the CSS file, and an initialisation script, in selected webpages, using a freely-available browser extension\footnote{ We used Code Injector:\\\url{https://addons.mozilla.org/en-US/firefox/addon/codeinjector/}}.
Moreover, by building a public repository of initialisation scripts (\eg one per domain), end-users would not even have to specify the selectors to use by themselves.
This could make menu adaptation on the web seamless and accessible to anyone, without requiring any technical knowledge.


\nopagebreak

\section{Conclusion and Outlook}
This paper has presented SAM, a JavaScript framework for developing and deploying adaptive menus on the web.
It allows to compose policies and styles to easily explore new types of adaptive techniques.
In the future, we aim to provide additional technical resources for inquiring user experience via studies which include adaptive menus supported by SAM.
Furthermore, we intend to extend the framework with more policies and styles, which better cover existing literature.

\begin{acks}
This work has been funded by the European Research Council (ERC) under the European Union's Horizon 2020 research and innovation programme (grant agreement 637991).
\end{acks}

\bibliographystyle{ACM-Reference-Format}
\bibliography{refs}

\end{document}